\pdfoutput=1
\documentclass[aps,prd,twocolumn,superscriptaddress, nofootinbib]{revtex4}

\usepackage[T1]{fontenc}
\usepackage{ae,aecompl}

\usepackage{abrev-jour-long}

\usepackage{hyperref}
\hypersetup{
    colorlinks,
    citecolor=blue,
    linkcolor=blue,
    urlcolor=blue
}

\usepackage{graphicx}	% Including figure files
\usepackage{amsmath}	% Advanced maths commands
\usepackage{amssymb}	% Extra maths symbols
\usepackage{float}
\usepackage{color}

\begin{document}

\title[Particle motion around luminous neutron stars: effects of deviation from Schwarzschild spacetime]{Particle motion around luminous neutron stars:\\ effects of deviation from Schwarzschild spacetime}

\author{Ronaldo S. S. Vieira}
\email{ronaldo.vieira@ufabc.edu.br}
\affiliation{Centro de Ci\^encias Naturais e Humanas, Universidade Federal do ABC, 09210-580 Santo Andr\'e, SP, Brazil}

\author{Maciek Wielgus}
\email{maciek@wielgus.info}
\affiliation{Max-Planck-Institut f\"ur Radioastronomie, Auf dem H\"ugel 69, D-53121 Bonn, Germany} 
\affiliation{Research Centre for Computational Physics and Data Processing, Institute of Physics, \\ Silesian University in Opava, Bezru\v{c}ovo n\'am.~13, CZ-746\,01 Opava, Czech Republic} 

% These dates will be filled out by the publisher
\date{Accepted XXX. Received YYY; in original form ZZZ}

% Enter the current year, for the copyright statements etc.
%\pubyear{2019}

% Don't change these lines

\begin{abstract}
We study trajectories of test particles around a luminous, static, spherically symmetric neutron star, under the combined influence of gravity and radiation. In general relativity, for Schwarzschild spacetime, an equilibrium sphere (the Eddington Capture Sphere) is formed for near-Eddington luminosities. We generalize these results to a broad class of static, spherical spacetimes. We also study the dynamics of particles in a strong radiation field in spherical spacetimes. The results are illustrated for two cases, Reissner-Nordstr\"{o}m spacetime of a charged spherical object in general relativity and Kehagias-Sfetsos spacetime, arising from the Ho\v{r}ava–Lifshitz gravity theory. Our findings apply to neutron stars under gravitational field equations different from the vacuum Einstein field equations of general relativity, such as in modified theories of gravity, the only requirement being that test particles follow geodesics in the absence of the radiation field. The effects that we describe are, in principle, measurable through observations of X-ray bursts of neutron stars. Hence, detailed future studies could use such observations to test gravity theories in the strong-field regime, provided that the impact of the spacetime geometry can be disentangled from the astrophysical uncertainties.

%We show that it is possible, in principle, to test the gravity theory by investigating such test particle equilibria and trajectories with observations of neutron stars' X-ray bursts.
\end{abstract}

\maketitle

%%%%%%%%%%%%%%%%%%%%%%%%%%%%%%%%%%%%%%%%%%%%%%%%%%

%%%%%%%%%%%%%%%%% BODY OF PAPER %%%%%%%%%%%%%%%%%%

\section{Introduction}
\label{sec:intro}

Radiation exerts force on test particles, influencing their trajectories and causing their deviation from timelike geodesics. For a luminous source the radiation force may be comparable to the effective gravity. In particular, the Eddington luminosity
\begin{equation}
\label{eq:LEdd}
L_{\rm{Edd}} = \frac{4\pi m GM c}{ \sigma_{\rm T}}
\end{equation}
results in a global equilibrium between the Newtonian gravity and the radiation pressure for a spherical source of mass $M$. In Eq.~(\ref{eq:LEdd}) we assume hydrogen plasma, hence $\sigma_{\rm T}$ is the Thomson cross section and $m$ is the proton mass. $G$ and $c$ denote the gravitational constant and speed of light, respectively.  The problem of the gravity-radiation equilibrium is more complicated for the relativistic case -- in general relativity (GR) equilibrium can only be established at a~particular radius, referred to as the Eddington Capture Sphere (ECS; see \cite{Stahl2012, Wielgus2012}). The ECS radius $r_{\rm ECS}$ depends on the stellar luminosity as seen by a distant observer $L_{\infty}$, and in Schwarzschild spacetime corresponds to
\begin{equation}
    r_{\rm ECS} = \frac{2\,GM/c^2}{1 - \lambda^2} \ ,
\label{eq:RECS}
\end{equation}
where $\lambda \equiv L_\infty/L_{\rm Edd}$. In Eq.~(\ref{eq:RECS}), $L_{\infty} = L_{\rm{Edd}}$ represents equilibrium established at infinity. The ECS may form as a transient phenomenon during type-I luminous X-ray bursts on neutron stars \cite{Tawara1984,Lewin1993}, particularly in cases exhibiting photospheric expansion \cite{Lewin1984,Strohmayer2006}, when the radiation originating from a thermonuclear explosion engulfing the neutron star surface temporarily balances the effective gravity. Studying such events may lead to constraints on neutron star equation of state \cite{Ozel2006, Lattimer2012}, mass, radius \cite{Galloway2008, Bollimpalli2019}, and spin \cite{Muno2001}. In this paper we consider whether the location of the equilibrium surface and test-particle dynamics are sensitive to deviations of the underlying metric from the Schwarzschild solution. While several authors considered deviations caused by the central mass rotation \cite{SokOh2010,Bini2011,DeFalco2019, Wielgus2019}, quadrupolar deviation of the metric from spherical symmetry \cite{DeFalco2020}, or both \cite{DeFalco2021}, we focus on the effects related to modifying the underlying gravity field equations (and/or modifications of the background metric).

Furthermore, the aim of this paper is to demonstrate the impact of the spacetime deviation from the canonical Schwarzschild case on properties of systems including luminous neutron stars, illustrating it with concrete examples. Such deviations in spacetime geometry could in principle be detected through detailed modeling of X-ray light curves of bursting neutron stars, offering a previously unexplored avenue to test theories of gravity in the strong field and large curvature regime. This task will, however, require careful modeling in order to disentangle the effects of astrophysical configuration, star rotation, and potential deviation from general relativity.

%==================================
%==================================
\section{Equations of motion}
\label{sec:eom}
%==================================
%==================================

%

%\cite{SokOh2010,Stahl2012,Stahl2013,Wielgus2019,DeFalco2019}
%\rv{some c factors are missing here and there...\newline}

We consider a static, spherically symmetric spacetime metric in the form
\begin{equation}
   {\rm d} s^2  = -\xi\, c^2 {\rm d} t^2 + \xi^{-1}\,{\rm d}r^2 + r^2 \,{\rm d} \theta^2  + r^2 \sin^2\theta \,{\rm d} \phi^2 \, ,
\label{eq:metric}
\end{equation}
which describes a wide class of spherically symmetric spacetimes,
where $\xi$ depends only on the radial coordinate $r$, $\xi \equiv \xi(r)$. Schwarzschild spacetime is recovered for $\xi(r) = 1 - 2M/r$, where $M$ is the geometric mass of the star. In the presence of a radiation flux $F^\mu$, the test-particle equation of motion reads
\begin{equation}
a^\mu = u^\nu \nabla_\nu u^\mu = \frac{\rm d^2 x^\mu}{\rm d \, {\tau}^2 \,} + \Gamma^\mu_{\ \nu \rho}u^\nu u^\rho = \frac{\sigma_{\rm T}}{m {c}} F^\mu \, ,
\label{eq:eom0}
\end{equation}
where $a^\mu$ is the particle's four-acceleration and $u^\mu$ is its four-velocity. Equation~(\ref{eq:eom0}) corresponds to the covariant formulation of Newton's second law with a standard electrodynamic radiation force term $f^\mu \propto F^\mu$, proportional to the radiation flux measured in the frame of the test particle. The only non-zero Christoffel symbols are

\begin{align}
   & \Gamma^r_{\ t t} = \frac{c^2 \xi \xi'}{2} \, , \\
   &  \Gamma^r_{\ r r} = -\frac{\xi'}{2\xi}  \, ,  \\
   &  \Gamma^r_{\ \theta \theta} = -r \xi \, , \\
   & \Gamma^r_{\ \phi \phi} = -r \xi \sin^2 \theta \, , \\
   & \Gamma^\phi_{\ r \phi} = \frac{1}{r} \, ,
\end{align}
%\begin{equation}
%\Gamma^r_{\ t t} = \frac{c^2 \xi \xi'}{2} , \
%    \Gamma^r_{\ r r} = -\frac{\xi'}{2\xi} , \ \ \Gamma^r_{\ \theta \theta} = -r \xi , \ 
%    \ \Gamma^r_{\ \phi \phi} = -r \xi \sin^2 \theta 
%\end{equation}
%and
%\begin{equation}
%\Gamma^\phi_{\ r \phi} = \frac{1}{r} \ ,
%\end{equation}
where we denote
\begin{equation}
\xi' \equiv \frac{{\rm d} \xi}{{\rm d} r}    \, .
\end{equation}
 Using the projection tensor $h^\mu_{\ \nu}$ onto the particle's local 3-space, Kronecker tensor $ \delta^\mu_{ \ \nu}$, and the radiation stress-energy tensor $T^{\nu \rho}$, the radiation flux $F^\mu$ measured in the particle's local rest frame can be represented as \footnote{Unless explicitly stated, we hereafter use geometrized units ($G=c=1$).}
\begin{equation}
\label{eq:force}
    F^\mu = h^\mu_{\  \nu} T^{\nu \rho}u_\rho = \left( \delta^\mu_{ \ \nu} + u^\mu u_\nu \right) T^{\nu \rho}u_\rho \, .
\end{equation}
 A fully covariant description of the radiation field of a static, uniformly radiating, spherical star in Schwarzschild spacetime in terms of the radiation tensor $T^{\nu \rho}$ was first given by \cite{Abramowicz1990}. This description can be readily generalized to a broader class of static spherical metrics. 
 The frequency-integrated specific intensity $I$ of radiation is uniform over all angles and given by 
\begin{equation}\label{eq:intensity}
I(r) = I(R) \left(\frac{\xi_{R}}{\xi} \right)^2\, , %= \frac{G M m c  }{\pi \sigma_{\rm T}} \frac{\xi_R}{\xi^2 R^2} \lambda ,
%\frac{L^\infty}{L_{\rm{Edd}}} \nonumber \\
%&= \frac{m M (1-\frac{2M}{R})}{\pi \sigma_{\rm T} R^2 (1-\frac{2M}{r})^2} \lambda \ ,
\end{equation}
 where $R$ is the stellar radius and $\xi_R \equiv \xi(R)$. The apparent viewing angle $\alpha_0$ of the star, measured at radius $r>R$, is
\begin{equation}
\sin \alpha_0 =   \frac{R}{r} \left(\frac{\xi}{\xi_{\rm R}} \right)^{1/2}  \, .
\label{eq:alpha0}
\end{equation}
Equation~(\ref{eq:alpha0}) naturally requires that $r > R$, with another requirement that $r > r_\gamma$, where $r_\gamma$ is the photon sphere radius, a solution to $\xi'r - 2 \xi = 0$, see e.g. \cite{Wielgus2021}. It is generally expected that $R> r_\gamma$ for neutron stars, while it may not necessarily be the case for hypothetical more compact strange stars, e.g. \cite{Haensel1986}.

\begin{figure}
\centering
\includegraphics[width=0.485\textwidth]{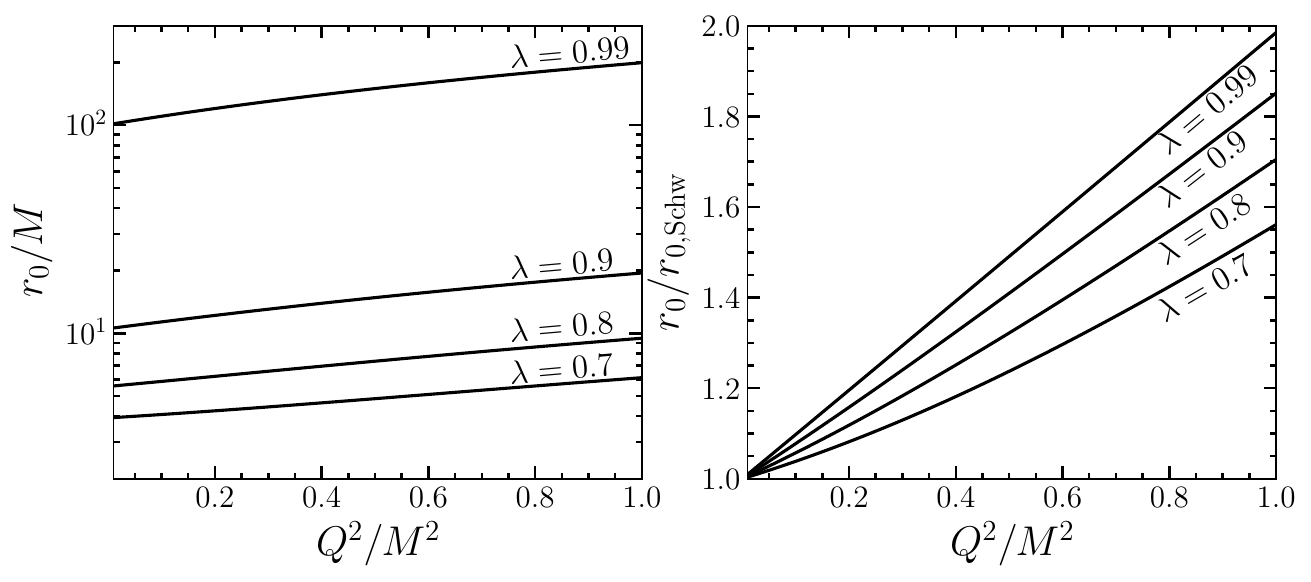}
\\ \hfill \\
\includegraphics[width=0.485\textwidth]{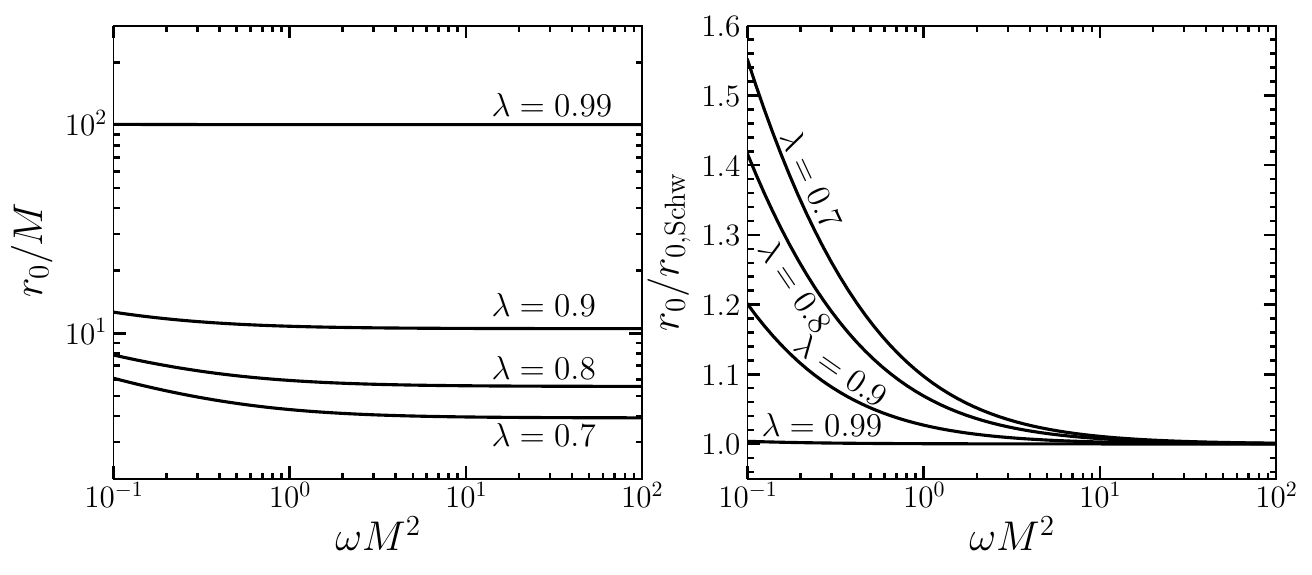}
\caption{Radius of the gravity-radiation equilibrium sphere for Reissner-Nordstr\"om (top row) and Kehagias-Sfetsos (bottom row) spacetimes for various luminosity parameters $\lambda$. \textit{Left column:} radius as function of the metric parameter $(Q/M)^2$ in RN spacetime  ($\omega M^2$ in KS spacetime), $Q=0$ ($\omega M^2 \rightarrow \infty$) corresponds to Schwarzschild spacetime. \textit{Right column:} relative change in the location of the equilibrium sphere with respect to the Schwarzschild case.}
\label{fig:RNKS_location}
\end{figure}

\begin{figure*}
\centering
\includegraphics[width=0.65\columnwidth]
{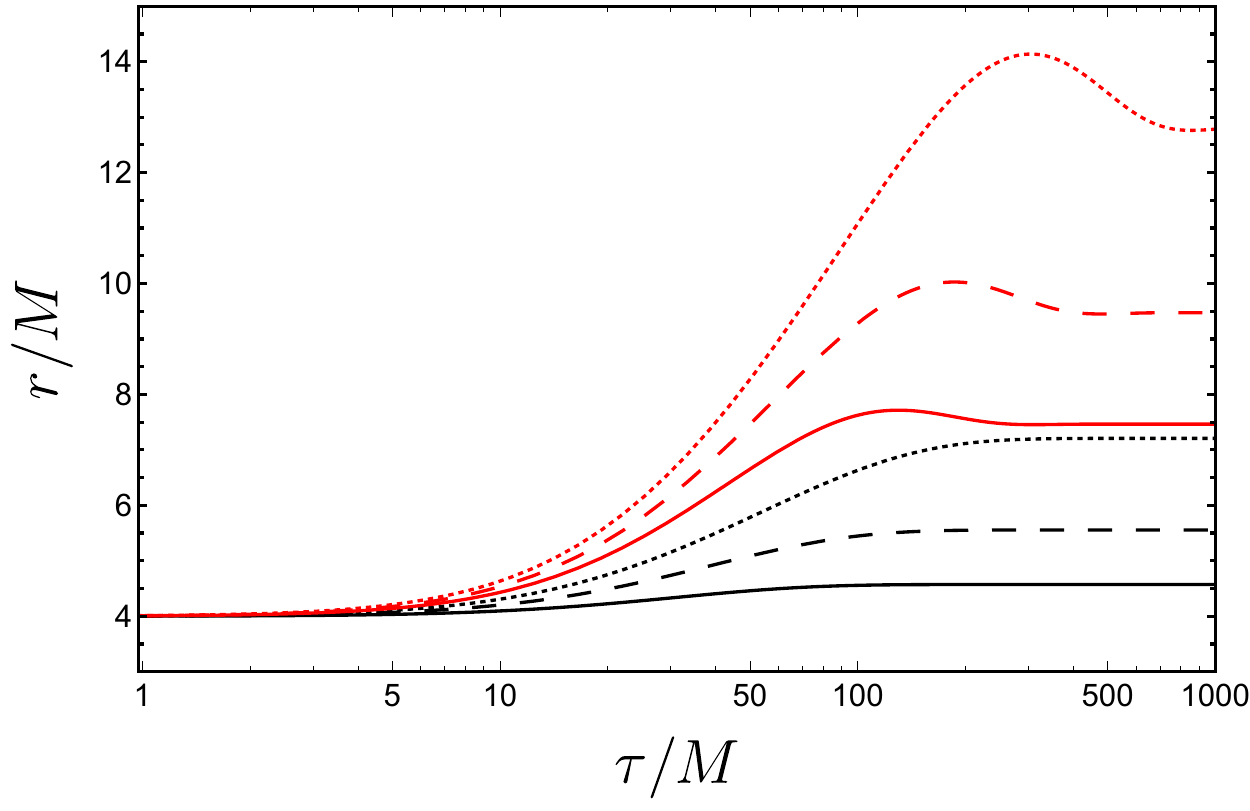}\quad
%\\ \hfill \\
\includegraphics[width=0.65\columnwidth]
{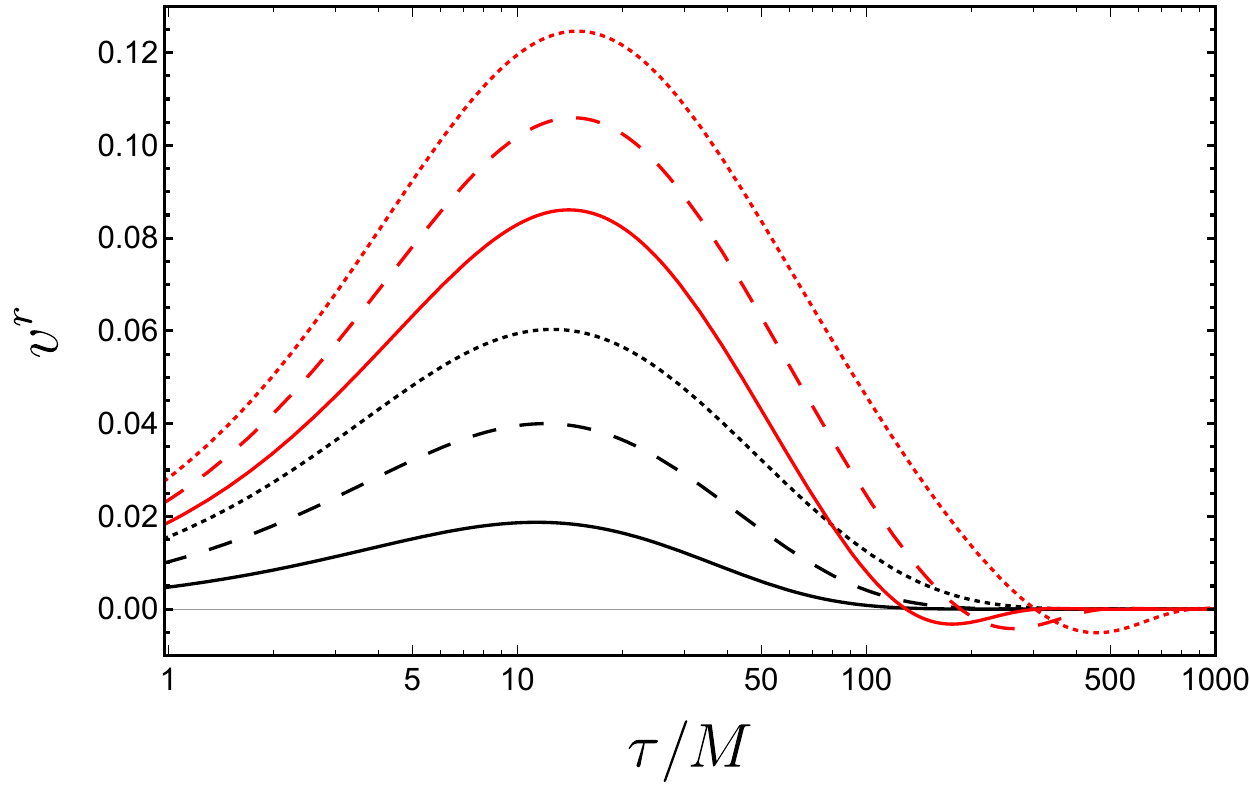}\quad
%\\ \hfill \\
\includegraphics[width=0.65\columnwidth]
{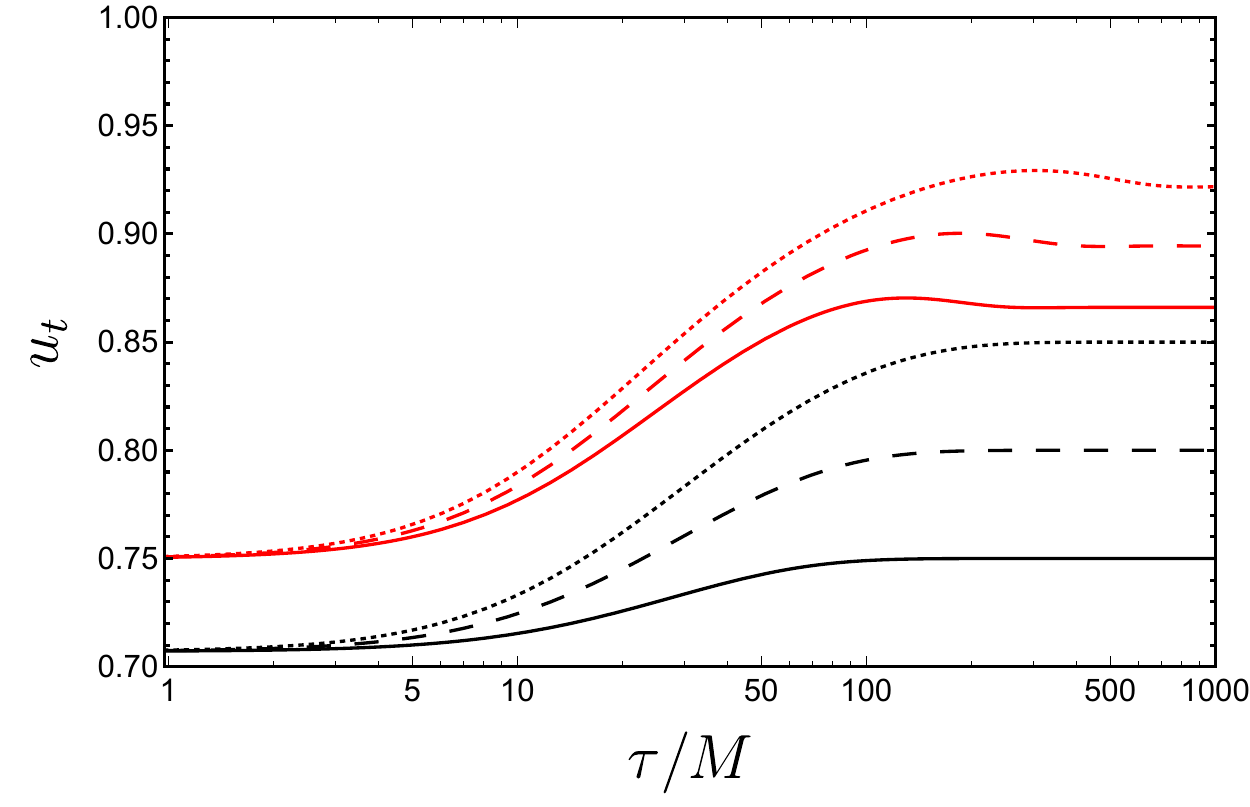}
\\ \hfill \\
\includegraphics[width=0.65\columnwidth]
{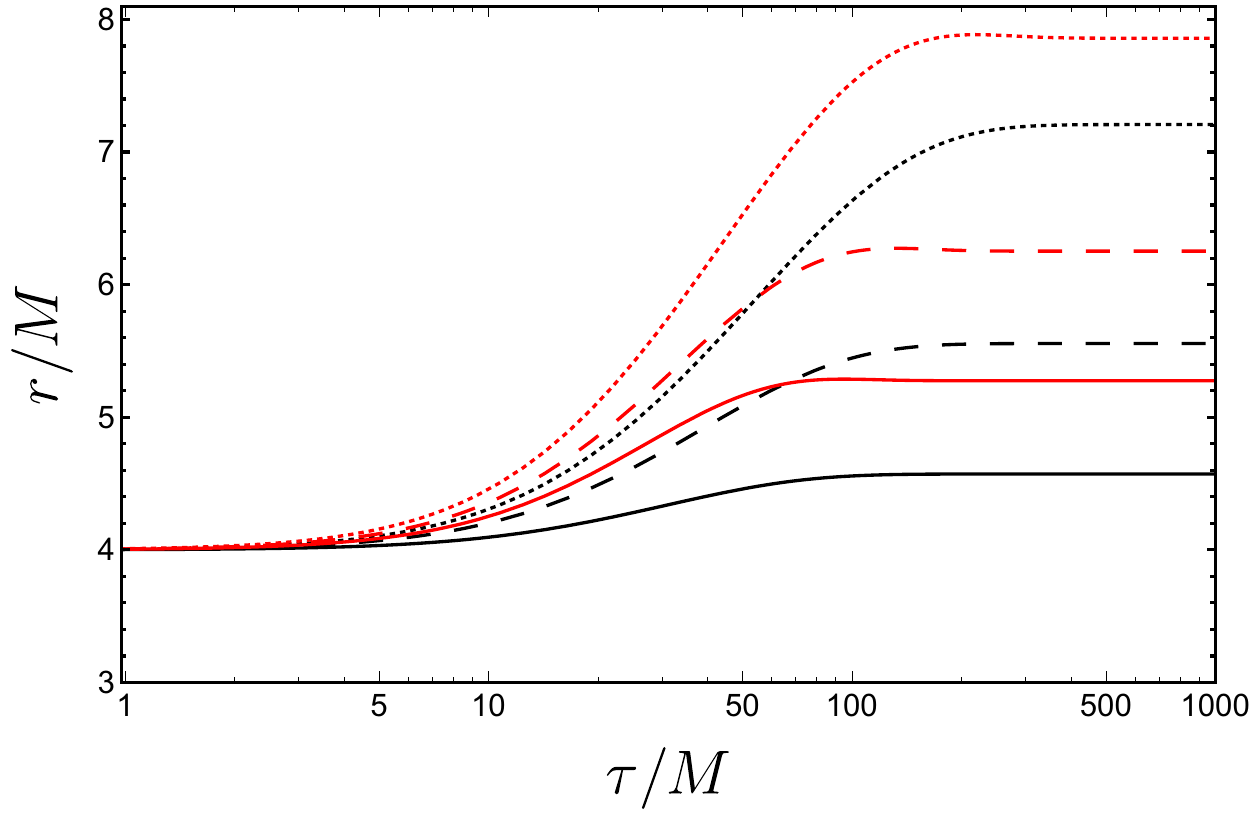}\quad
%\\ \hfill \\
\includegraphics[width=0.65\columnwidth]
{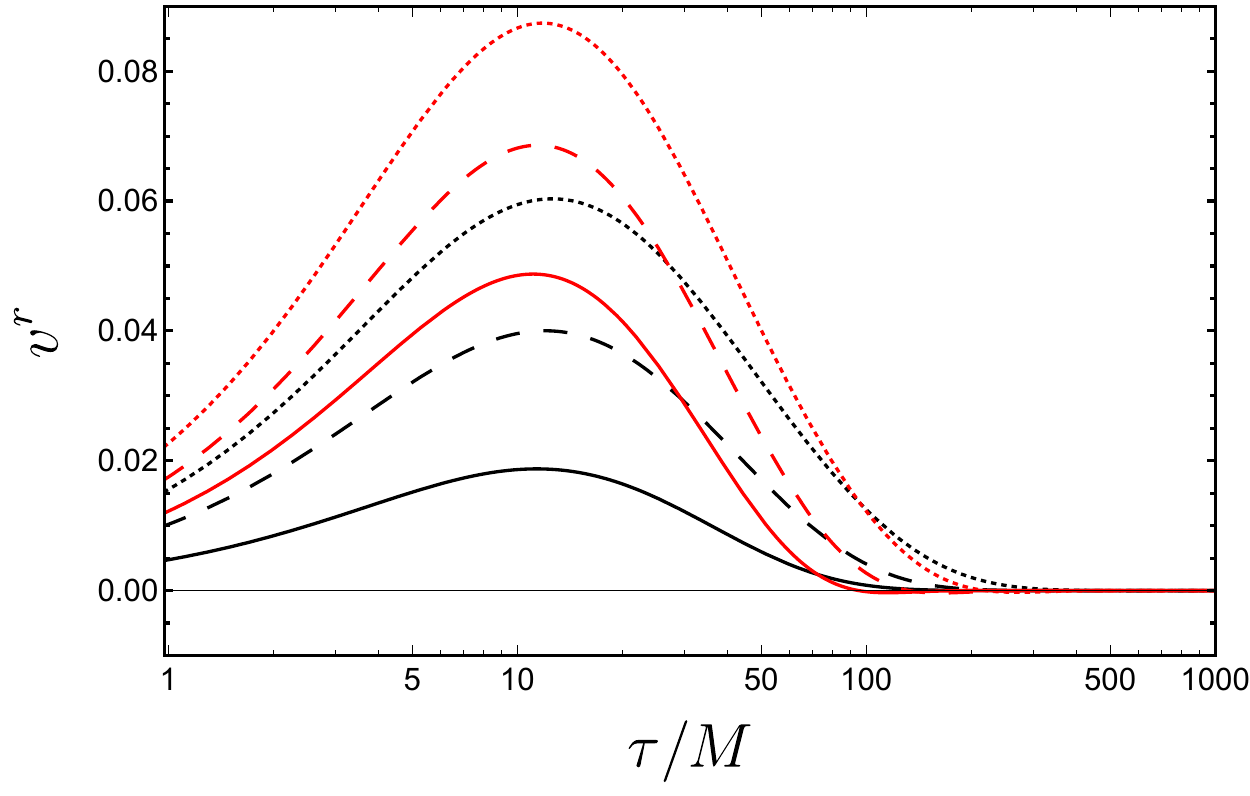}\quad
%\\ \hfill \\
\includegraphics[width=0.65\columnwidth]
{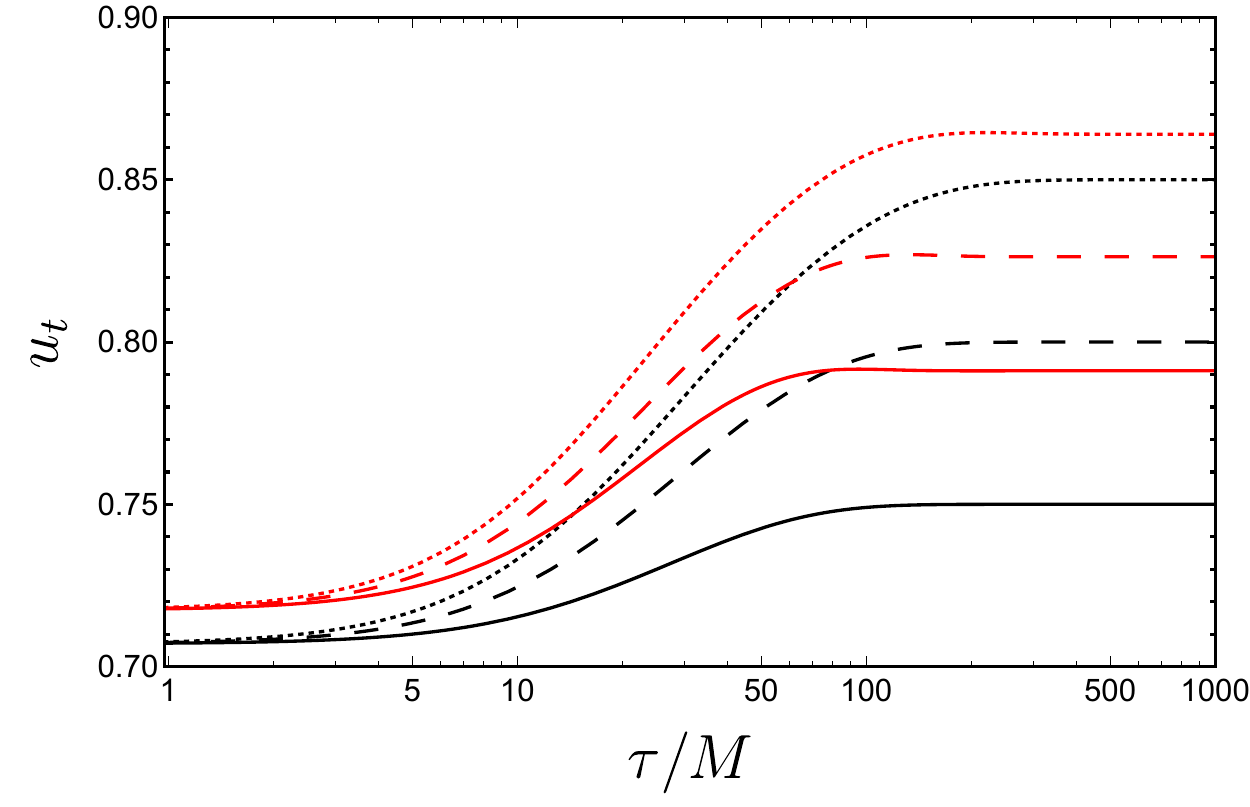}
\caption{Radius (left column), radial velocity component $v^r = \sqrt{-g_{rr}/g_{tt}}\,u^r/u^t$ (middle column), and energy $-u_t$ (right column) as functions of the proper time $\tau$ for particles ejected from the neutron star surface located at $R = 4M$ by the instantaneous burst of radiation with the luminosity parameter $\lambda=0.75$ (solid), $\lambda=0.80$ (dashed), and $\lambda=0.85$ (dotted). The time unit corresponds to $M \approx 10^{-5}$\,s for a neutron star mass of $2 \,M_\odot$. The black curves correspond to Schwarzschild spacetime. The red curves in the top row correspond to Reissner-Nordstr\"om spacetime with $(Q/M)^2 = 1$. The red curves in the bottom row correspond to Kehagias-Sfetsos spacetime with $\omega M^2=0.5$.}
\label{fig:RNKS_quantities}
\end{figure*}

The radiation stress-energy tensor is given by $T^{\mu\nu} = I(r)\,A^{\mu\nu}$, where the dimensionless components of the $A^{\mu\nu}$ tensor take the following form in the orthonormal tetrad associated with the coordinate basis vectors,
\begin{align}
    \label{eq:Ttt}
    A^{\hat{t} \hat{t}} &= 2 \pi  \left( 1 - \cos \alpha_0 \right) \, , \\
    A^{\hat{t} \hat{r}} &= \pi  \sin^2 \alpha_0 \, , \\
    A^{\hat{r} \hat{r}} &=  \frac{2 \pi}{3}  \left(1 - \cos^3 \alpha_0 \right) \, , \\
    A^{\hat{\phi} \hat{\phi}} &= A^{\hat{\theta} \hat{\theta}} =  \frac{\pi }{3}  \left(\cos^3 \alpha_0 - 3 \cos \alpha_0 + 2\right) \, .
\end{align}
Tetrad frame components $A^{\hat{\mu} \hat{\nu}}$ are related to the coordinate frame components $A^{\mu \nu}$ through
\begin{equation}
    A^{\hat{\mu} \hat{\nu}} = A^{\alpha\beta}e_{\ \alpha}^{\hat{\mu}} e_{\ \beta}^{\hat{\nu}}\, ,
\end{equation}
where tetrad components in the coordinate basis are given by 
\begin{equation}
    e_{\ \mu}^{\hat{\mu}} = |g_{\mu \mu}|^{1/2}\, .
\end{equation}
We parameterize the radiation strength in terms of the dimensionless parameter $\lambda$
\begin{align}
\lambda = \frac{L_{\infty}}{L_{\rm Edd}} = \frac{\xi_R L(R)}{L_{\rm Edd}} &= \frac{ 4 \pi^2 \xi_R R^2 I(R)}{L_{\rm Edd}}  \nonumber \\
&= \frac{\pi \sigma_{\rm T} \xi_R R^2 I(R) }{G M m c}
\label{eq:lambda}
\end{align}
where $M$ in a general spacetime corresponds to the Arnowitt-Deser-Misner mass, so we assume that the spacetime approaches Schwarzschild solution in the weak field limit (large radius). Putting together Eqs.~(\ref{eq:eom0})--(\ref{eq:lambda}) and leveraging spherical symmetry to choose a coordinate system with $\theta = \pi/2$, ${\rm d} \theta / {\rm d} \tau = {\rm d}^2 \theta / {\rm d} \tau^2 = 0 $, we obtain the full equations of motion for a test particle
\begin{eqnarray}
    \frac{{\rm d}^2 r}{{\rm d} \tau^2} &=& - \frac{1}{2}\xi' + r\,\left(\xi - \frac{r\xi'}{2}\right) \bigg(\frac{{\rm d}\phi}{{\rm d}\tau}\bigg)^2     \label{eq:eom_dr}\\
   && + \frac{\lambda M\,\xi_R}{\pi R^2\xi^2}\,\left[\xi\, A^{\hat{t} \hat{r}} u^t - \left(A^{\hat{r} \hat{r}} + \chi\right) \frac{{\rm d}r}{{\rm d} \tau}\right] \, , \nonumber\\
    \frac{{\rm d}^2 \phi}{{\rm d}\tau^2} &=& - \left[ \frac{1}{r}\frac{ {\rm d} r}{{\rm d} \tau} + \frac{\lambda M\,\xi_R}{\pi R^2\xi^2}\,\left(A^{\hat{\phi} \hat{\phi}} + \chi\right) \right]\frac{{\rm d}\phi}{{\rm d}\tau}  \label{eq:eom_dphi} \, , 
\end{eqnarray}
where $\tau$ is the particle's proper time and we denote
\begin{eqnarray}
    \chi \equiv& & \xi\, A^{\hat{t} \hat{t}}  (u^t)^2 + \xi^{-1}A^{\hat{r} \hat{r}} \left(\frac{{\rm d} r}{{\rm d} \tau}\right)^2 \nonumber\\
     & & +\, r^2\,A^{\hat{\phi} \hat{\phi}} \left(\frac{{\rm d} \phi}{{\rm d} \tau}\right)^2 - 2 A^{\hat{t} \hat{r}}u^t\, \frac{{\rm d} r}{ {\rm d} \tau}\, .
    \label{eq:eomEND}
\end{eqnarray}
Additionally, from the four-velocity normalization $u^\alpha u_\alpha = -1$ we have
\begin{equation}
    u^t = \xi^{-1/2}\bigg[\, 1 + \xi^{-1}\Big(\frac{{\rm d} r}{ {\rm d} \tau}\Big)^2 + r^2\,\Big(\frac{{\rm d} \phi}{{\rm d} \tau}\Big)^2\bigg]^{1/2} \, .
\end{equation}

There are four components in Eq.~(\ref{eq:eom_dr}), that can be associated with gravitational acceleration $\propto \xi'$, centrifugal acceleration $\propto ({\rm d} \phi/ {\rm d} \tau)^2$, radiation pressure $\propto \lambda\, u^t$, and radial radiation drag $\propto \lambda\, {\rm d} r/ {\rm d}\tau$. In Eq.~(\ref{eq:eom_dphi}) we recognize the Coriolis acceleration $\propto ({\rm d} r / {\rm d} \tau)({\rm d} \phi/ {\rm d} \tau)$. The second component of Eq.~(\ref{eq:eom_dphi}) is the Poynting–Robertson drag force, removing the particle's angular momentum through interaction with the radiation field. In general relativity these effects were studied in, e.g., \cite{Bini2009,Stahl2013,Mishra2014,DeFalco2019,Fragile2020}. Equations~(\ref{eq:eom_dr})--(\ref{eq:eomEND}) reduce to the ones presented in \cite{Stahl2012} for the case of Schwarzschild spacetime.

\subsection{Back-reaction on spacetime}
Another interesting question that one may ask is whether the strong radiation field, balancing the effective gravity, should be included in the calculation of the spacetime metric from the underlying field equations as a~non-zero stress-energy tensor component. While a complete, rigorous and non-linear treatment of this problem is complicated even for fixed underlying field equations, one can make a rather simple order-of-magnitude argument following Section 4 of \citep{Wielgus2014}, where a very similar problem of radiation field back-reaction on the Schwarzschild spacetime geometry was considered. Assuming Einstein vacuum field equations we characterize the magnitude of the Ricci part of the Riemann tensor by computing a~scalar quantity
\begin{equation}
\mathcal{R} = R^{\mu \nu}R_{\mu \nu} = \kappa^2 T^{\mu \nu}T_{\mu \nu} \propto \frac{(\lambda L_{\rm Edd})^2}{r^4}
\end{equation}
with $\kappa = 8 \pi G/ c^4$. After some amount of algebra and taking liberal overestimates of various quantities such as stellar radius $R$ or $\xi_R$ we conclude that
\begin{equation}
\mathcal{R} < 10^{-40} \frac{c^8}{G^4 M^4}
\end{equation}
near the surface of the Eddington-luminosity neutron star. This value characterizes the impact of the stress-energy tensor $T^{\mu \nu}$ on the spacetime geometry and can be compared with the Weyl part $C^{abcd}$ of the Riemann tensor $R^{abcd}$, characterizing the magnitude of the spacetime curvature in the absence of $T^{\mu \nu}$. The latter can be evaluated for the vacuum background metric simply with the Kretschmann scalar, which characterizes the Weyl tensor magnitude for a vacuum background spacetime
%\begin{equation}
%\mathcal{K} = C^{abcd} C_{abcd} = \frac{48}{r^6} \frac{G^2 M^2}{c^4} > %10^{-11} \frac{c^8}{G^4 M^4} ,
%\end{equation}
\begin{align}
\mathcal{K} &= R^{abcd} R_{abcd} = C^{abcd} C_{abcd} \nonumber \\
&= \frac{48}{r^6} \frac{G^2 M^2}{c^4} > 10^{-11} \frac{c^8}{G^4 M^4}
\end{align}
where we only assumed that $r < 100\,GM/c^2$. While this example is limited to Schwarzschild background and only constitutes a linear approximation, as the stress-energy tensor is evaluated in the background spacetime without accounting for the back-reaction, the huge gap between the magnitudes of $\mathcal{K}$ and $\mathcal{R}$ shows that the back-reaction is negligible for the Eddington-luminosity radiation around a neutron star. As a matter of fact, it would take over 10 orders of magnitude increase of the stellar luminosity for the energy of radiation to start contributing to the spacetime curvature in an appreciable way, as we show
that $\mathcal{R}/\mathcal{K} < 10^{-29}$
%\begin{equation}
%\frac{\mathcal{R}}{\mathcal{K}} < 10^{-29}
%\end{equation}
and $\mathcal{R} \propto \lambda^2$.

%==========================
\subsection{Radial equilibrium surface}
%==========================

For a static test particle we have
%\rv{(for later: $T^{\hat{t} \hat{r}} = c T^{{t}{r}}$)\newline}
$u^\mu = \delta^\mu_{\ t} u^t$ and $u^t = \xi^{-1/2}$, %$u_t = -\xi^{1/2}$, \rv{better $u^t = \xi^{-1/2}$?} 
hence the only non-zero components in Eq.~(\ref{eq:eom_dr}) correspond to gravitational acceleration and radiation pressure and Eq.~(\ref{eq:eom_dphi}) is trivially fulfilled. The whole system of equations of motion simplifies to
\begin{equation}
    0 = - \frac{1}{2}\xi' + \frac{\lambda M}{r^2 \xi^{1/2}}
\label{eq:static_forces}
\end{equation}
for $r>r_\gamma$ and at the equilibrium radius we have
\begin{equation}
2 M \lambda = r^2 \xi^{1/2} \xi'\, .
\label{eq:ECS_general}
\end{equation}
If Eq.~(\ref{eq:ECS_general}) has a solution for some $r = r_0 > R$, then it physically corresponds to the gravity-radiation equilibrium surface suspended above the luminous stellar surface. For Schwarzschild metric Eq.~(\ref{eq:ECS_general}) reduces to $\lambda = \xi^{1/2}$, and hence the ECS formula given in Eq.~(\ref{eq:RECS}) is recovered. We denote this general equilibrium radius by $r_0$, and the Schwarzschild value $r_{\rm ECS} \equiv r_{0, \rm{Schw}}$.

To discuss the stability of the equilibrium, let us consider a~restoring force
\begin{equation}
 f(r) =  \frac{1}{2}\xi' - \frac{\lambda M}{r^2 \xi^{1/2}}   \, ,
\label{eq:ECS_force}
\end{equation}
with $f(r_0) = 0$. In order for the radial equilibrium at $r_0$ to be stable, we need 
\begin{equation}
    f'(r) = \frac{1}{2} \xi'' + \frac{\lambda M}{2r^2 \xi^{1/2}} \left(\frac{\xi'}{\xi} + \frac{4}{r}\right) \ 
\end{equation}
to be positive at $r_0$. Using the $f(r_0) = 0$ condition we find the stability condition at the equilibrium radius
\begin{equation}\label{eq:fprime}
    f'(r_0) = \frac{1}{2} \xi'' + \frac{1}{4} \xi' \left(\frac{\xi'}{\xi} + \frac{4}{r_0}\right) > 0 \ .
\end{equation}
Stability of gravity-radiation equilibria is thus a spacetime property, and for Schwarzschild spacetime equilibria are radially stable for all $r_0> r_\gamma$ \cite{Abramowicz1990, Stahl2012}. For the cases discussed further in this paper we always find stability for $r_0 > r_\gamma$.

%%%%%%%%%%%%%%%%%%%%%%%%%%%%%%%%%%%%%%%%%%%

\section{Metric examples}

We exemplify the general calculations given in Section~\ref{sec:eom} by considering two particular spacetime models: Reissner-Nordstr\"{o}m (RN) and Kehagias-Sfetsos \cite[KS; ][]{kehagiasSfetsos2009PhLB}. The former originates from the electrovacuum Einstein field equations and corresponds to the spacetime of a static electrically charged black hole (or exterior of a~charged spherical star). Recently a lot of interest in RN spacetime came from the brane cosmology \cite{Randall1999}, where the RN model is a solution of the effective field equations and the ``tidal charge'' corresponds to gravitational effects from the fifth dimension \cite{Dadhich2000}. The KS spacetime arises as a static, spherical solution representing a~black hole (or the exterior of a star) in Horava's gravity, a popular model of a~quantum field theory of gravity \cite{Horava2009}. The geodesic structure of the KS spacetime has been studied extensively by \cite{katkaVieiraEtal2015GRG, stuchlikSchee2014CQG, vieiraMarekEtal2014PRD}, for instance.

\begin{figure}
\centering
\includegraphics[width=0.99\columnwidth]{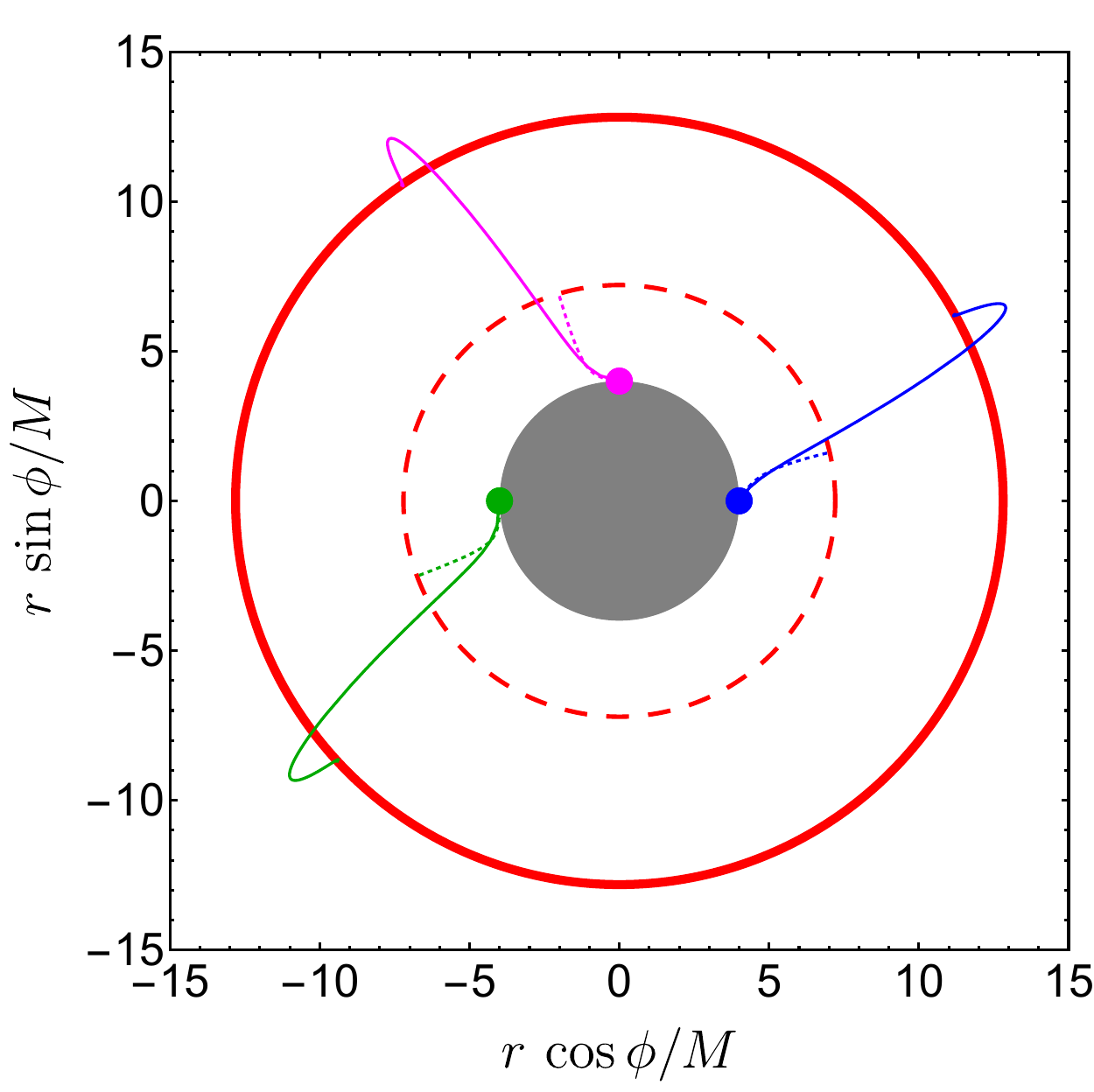}
\\ \hfill \\
\includegraphics[width=0.95\columnwidth]{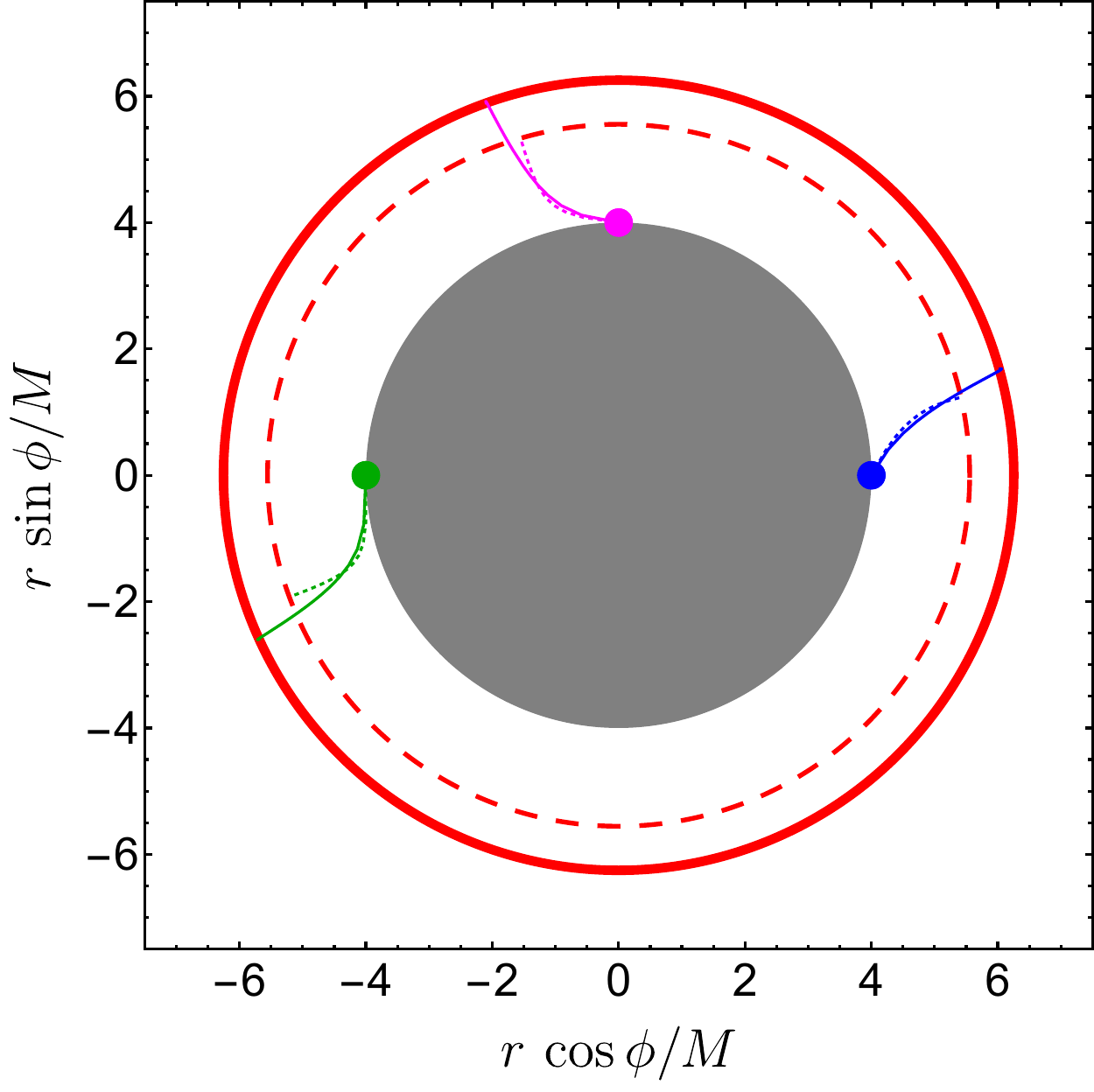}
\caption{Trajectories of particles ejected from the surface of a~neutron star in Reissner-Nordstr\"om (top, with $(Q/M)^2 = 1$) and Kehagias-Sfetsos (bottom, with $\omega M^2 = 0.5$) spacetimes. We assume a stellar radius of $R=4M$ (gray disc) and a~luminosity parameter $\lambda = 0.85$. The continuous red circles represent the corresponding equilibrium spheres, while the dashed red circles correspond to the Schwarzschild equilibrium (ECS). Each pair of colored trajectories corresponds to a particle ejected with a different initial velocity with respect to a local static observer at the stellar surface (starting at the marked dots): $v^r = 0.15$, $v^\phi = 0.2$ (blue); $v^r = 0$, $v^\phi = 0.4$ (green); $v^r = 0.05$, $v^\phi = 0.3$ (magenta). Continuous curves correspond to trajectories in RN and KS spacetimes, respectively; dashed curves represent trajectories in Schwarzschild spacetime computed for the same initial conditions.}
\label{fig:orbitsRNKS}
\end{figure}

%================================
%================================
\subsection{Equilibrium spheres}
%================================
%================================

In RN spacetime we have
\begin{equation}\label{eq:xiRN}
    \xi_{\rm RN}(r) = 1 - \frac{2M}{r} + \frac{Q^2}{r^2}\, ,
\end{equation}
where $Q$ is the star's charge, and the radial equilibrium Eq.~(\ref{eq:ECS_general}) gives a 4th order polynomial equation
%\begin{equation}
%    M^2(1-\lambda^2)r^4 - 2M(Q^2+M^2)r^3 + Q^2(Q^2 + 5M^2)r^2 - 4Q^4Mr + Q^6 = 0 .
%\end{equation}

\begin{align}
 0 =& \,M^2(1-\lambda^2)r^4 - 2M(Q^2+M^2)r^3  \nonumber \\ 
 &+ Q^2(Q^2 + 5M^2)r^2 - 4Q^4Mr + Q^6 \, .
\end{align}

The polynomial has a single real positive root $r_0$ such that $r_0 > M$, corresponding to the equlibrium sphere. The value of this equilibrium radius $r_0$ as a function of charge and luminosity parameter $\lambda$ is explored in Fig.~\ref{fig:RNKS_location} (top row). We see from the Figure that for all considered values of $\lambda$, we have $r_0/r_{0,\rm Schw}>1$.
%, in accordance with the fact that $\xi$ of Eq.~(\ref{eq:xiRN}) satisfies (\ref{eq:rECSgreater}). 
The equilibrium sphere radius generally grows with the charge $Q$ and can significantly deviate from the Schwarzschild solution for large values of $(Q/M)$ and large luminosity.

In KS spacetime we have 
\begin{equation}\label{eq:xiKS}
    \xi_{\rm KS}(r) = 1 +  \omega r^2 \left[1 - \left( 1 + \frac{4M}{\omega r^3} \right)^{1/2} \right] \, .
\end{equation}
The parameter $\omega$ is such that for $\omega\to\infty$ we recover Schwarzschild spacetime. Here we evaluate the location of the equilibrium sphere by solving Eq.~(\ref{eq:ECS_general}) numerically, and the results are presented in Fig.~\ref{fig:RNKS_location} (bottom row). The equilibrium sphere radius is once again larger than for the Schwarzschild case, but it is actually the lower neutron star luminosity (equilibrium forming for smaller $r_0$) resulting in a larger fractional deviation.

\begin{figure*}
\centering
\includegraphics[width=0.62\columnwidth]
{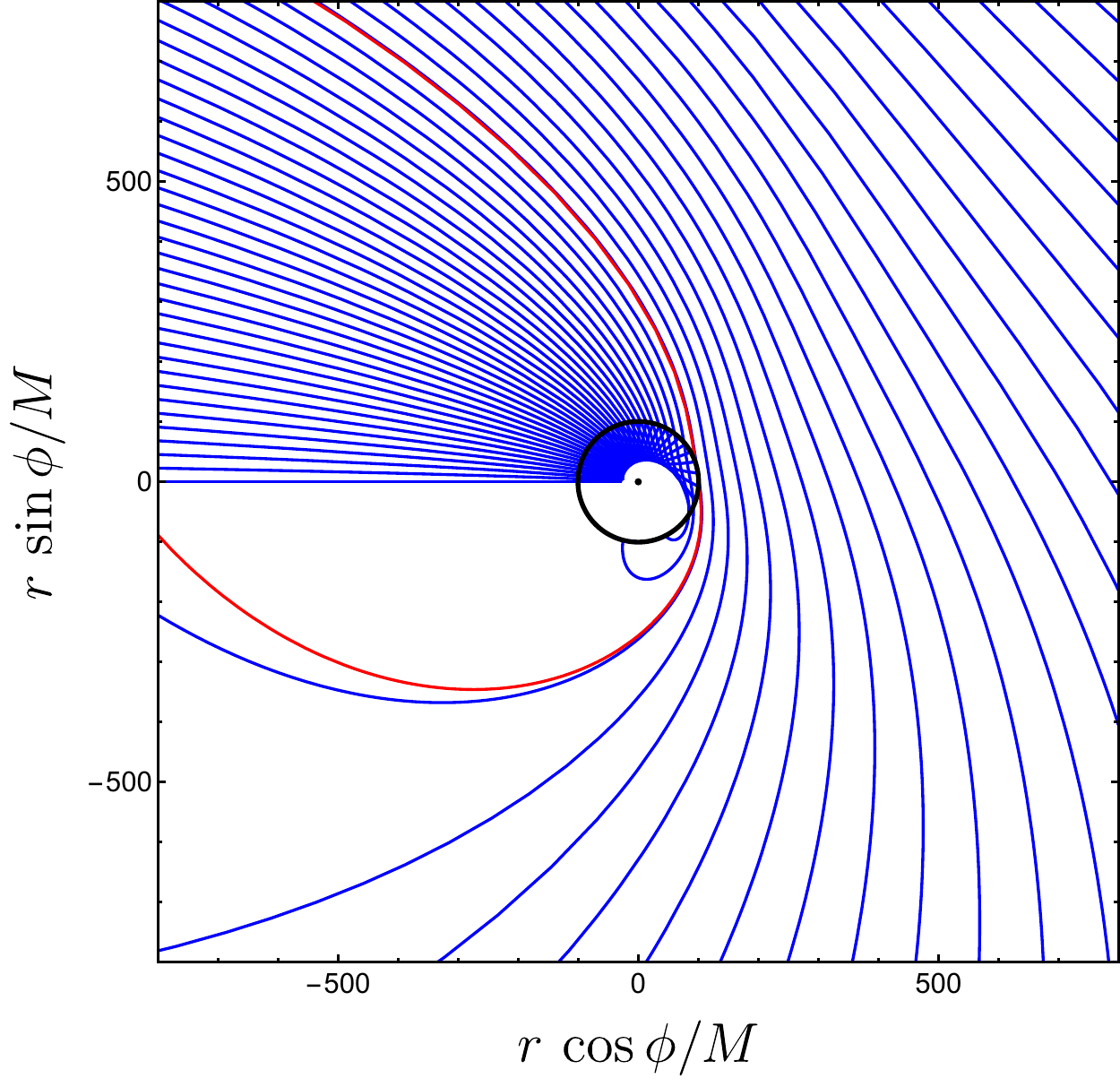}
\qquad
\includegraphics[width=0.62\columnwidth]
{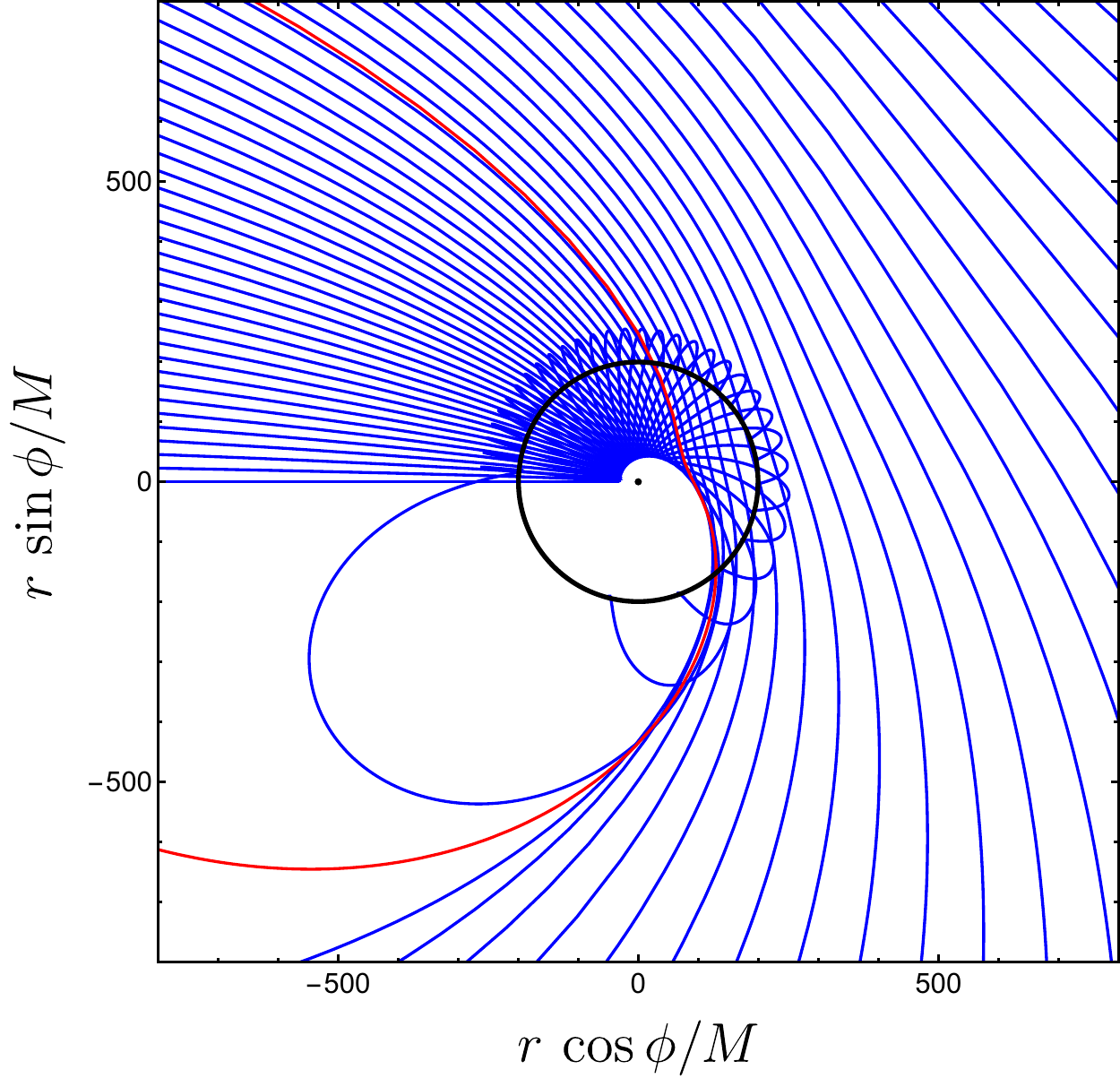}\qquad
 \includegraphics[width=0.62\columnwidth]
{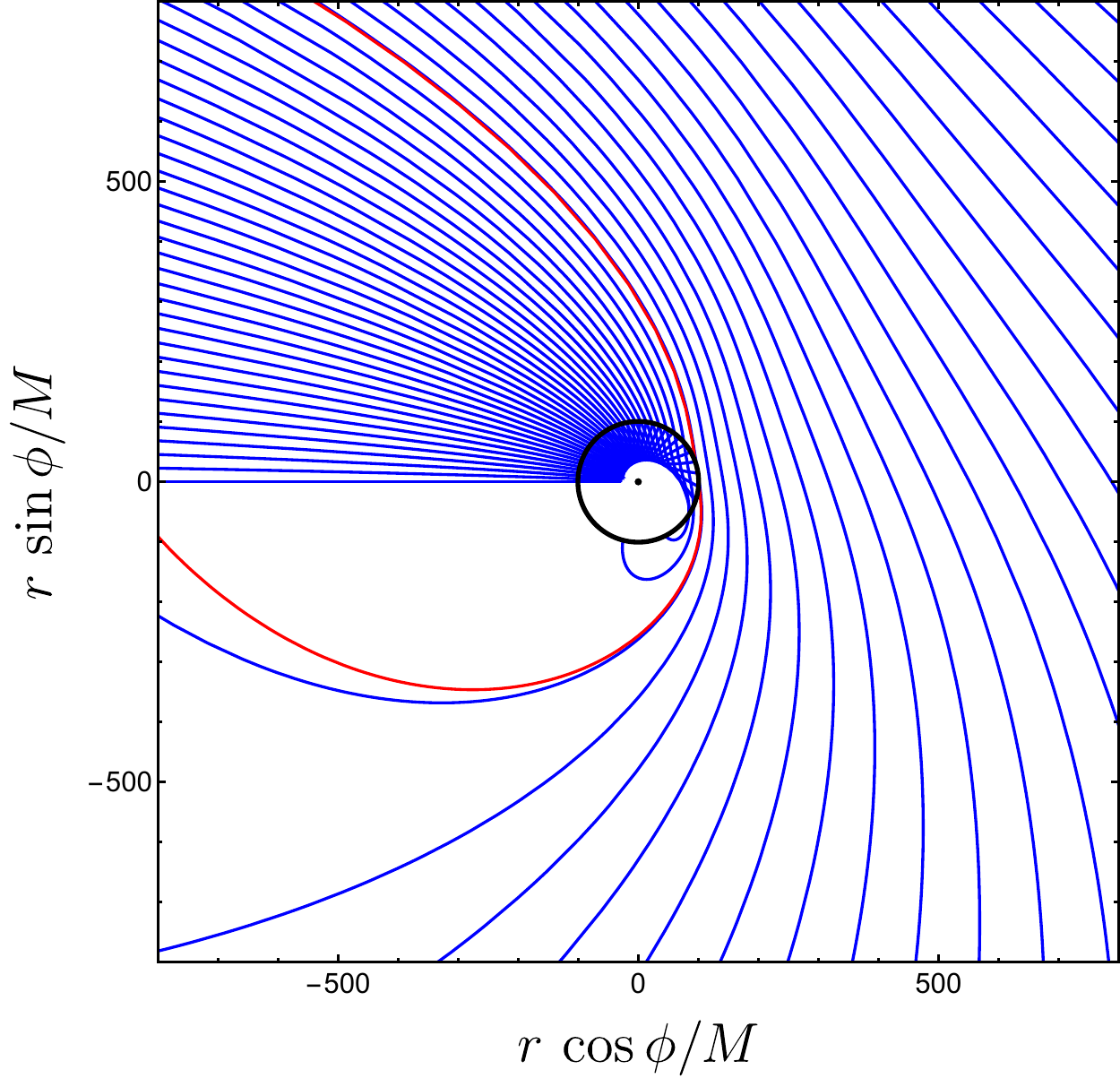}
\caption{Hoyle-Lyttleton accretion onto a luminous star. The particles are initiated at horizontal axis coordinate $x = -5000 M$, where the gravitational and radiative impact of the star is negligible, all with initial velocity of $v= 0.1\,c$ along the $x$ direction. The star is characterized by $R=4M$ and $\lambda = 0.99$. \textit{Left:} Schwarzschild spacetime. Trajectories with relatively low impact parameter are captured by the equilibrium sphere, whereas trajectories with high impact parameter are scattered (to infinity). The red curve corresponds to the lowest impact parameter for scattered trajectories (to infinity) considering a $1M$ resolution in the grid of impact parameters, $b_{\rm crit}=1697M$. \textit{Middle:} RN spacetime with $(Q/M)^2=1$. The red curve curve corresponds to the lowest impact parameter for scattered trajectories $b_{\rm crit}=1615M$. Typical captured trajectories in the middle panel penetrate the equilibrium sphere and then bounce, crossing it outwards and finally settling into it. The same qualitative picture happens in the right panel (KS spacetime with $\omega M^2 = 0.5$), but with a much smaller bouncing amplitude and with $b_{\rm crit}=1697M$, resembling the behavior in Schwarzschild spacetime.}
\label{fig:RNKS_HLyt}
\end{figure*}

%===============================
%===============================
\subsection{Particle ejection}
We study the scenario of a luminous outburst, expelling particles from the neutron star surface. Under the assumption of spherical symmetry, this becomes a~one-dimensional problem for the particle's radial motion. Similar problems were considered for the Schwarzschild case, e.g., by \cite{Wielgus2012,Stahl2013,Mishra2014}, including more general configurations. Here, we consider only the most physically relevant case, corresponding to particles ejected (with zero initial velocity) from the neutron star surface by a sudden burst of luminosity to a near-Eddington value. This is essentially a toy model for the photospheric radius expansion X-ray bursts of neutron stars \cite{Lewin1984, Lewin1993, Kuulkers2003}. Example results are shown in Fig.~\ref{fig:RNKS_quantities}. The timescale for the particles to settle on the equilibrium sphere is no longer than 1000\,$M$ for the chosen peak luminosity $\lambda$, which is of order of 0.01\,s for a typical neutron star. The radial velocity of ejected particles may be significantly larger for non-Schwarzschild spacetimes, reaching around 0.1\,$c$ for RN spacetime, before the particle settles on the equilibrium sphere. The kinetic energy of the particle is removed by the radiative drag force (Fig.~\ref{fig:RNKS_quantities}, second column), but the particle energy $(-u_t)$ is not conserved and grows as a result of the radiation pressure (Fig.~\ref{fig:RNKS_quantities}, last column).

In Fig.~\ref{fig:orbitsRNKS} we present orbits of particles ejected from the stellar surface with a sudden blast of luminosity for RN and KS spacetimes, with nonzero initial velocity components with respect to the star, comparing them with results for Schwarzschild spacetime obtained for the same initial condition. The figure demonstrates the efficiency of the angular momentum removal through radiation Poynting-Robertson drag force. The larger variation, related to detailed balance of gravity, radiation pressure and radiative drag forces, can be observed once again, particularly in the RN trajectories.

%===============================
%===============================
\subsection{Hoyle-Lyttleton accretion}
%===============================
%===============================

We consider a scenario of a luminous star passing through a cloud of matter with a fixed velocity, a Hoyle-Lyttleton accretion problem \cite{Hoyle1939}. This simple setup allows us to systematically explore the problem of the dynamics of particles with non-zero angular momentum with respect to the neutron star and the impact of the spacetime geometry.

In Fig.~\ref{fig:RNKS_HLyt} we present trajectories of particles in the Hoyle-Lyttleton scenario for Schwarzschild (left), extreme RN (middle), and extreme KS (right) spacetimes. The setup corresponds to a collection of particles approaching the neutron star from the left with an initial velocity of 0.1\,$c$, set beyond the gravitational sphere of influence of the neutron star. The particles are parameterized by their impact parameter, proportional to their (initial) specific angular momentum with respect to the neutron star. Particles with impact parameter $b$ lower than certain critical value $b_{\rm crit}$ (specific angular momentum $\ell <\ell_{\rm crit}$) are captured by the equilibrium sphere, hence the name Eddington Capture Sphere. The value of the $b_{\rm crit}$ parameter (and therefore of $\ell_{\rm crit}$) depends on the initial velocity, and also (weakly) on the neutron star diameter, impacting dynamical radiation drag forces and hence the efficiency of the angular momentum removal. Thus, the rate of mass capture, proportional to the capture cross section $\pi b_{\rm crit}^2$ of the luminous star, particle velocity, and particle mass density, is also sensitive to the spacetime geometry. These effects were investigated by, among others, \cite{Bini2009,Stahl2012,Wielgus2012} for the Schwarzschild spacetime case and by \cite{SokOh2011,Wielgus2019} for a~slowly rotating neutron star. Here we demonstrate that the role of the spacetime geometry may also be significant, which is particularly clear for the (extreme) RN case in Fig.~\ref{fig:RNKS_HLyt}. It is interesting to observe that (extreme) RN spacetime, with the equilibrium sphere larger than those of Schwarzschild and (extreme) KS cases, actually corresponds to a smaller cross section for capturing particles. This highlights the dynamic character of this problem, and the role of the relativistic radiation drag forces acting on a moving particle and depending on its four-velocity.

Luminous type-I neutron star bursts are transient phenomena of thermonuclear origin, fueled by the ignited material accreting onto the stellar surface. The bursts occur on timescales much longer than the dynamical ones (since $GM_{\odot}/c^3 \approx 5 \times 10^{-6}$\,s, the dynamical timescales involved are of order of milliseconds) and near-Eddington luminosity can be sustained for a~duration $\Delta T$ of tens of seconds \citep{Lewin1993}, while the ignited fuel reservoir burns away. Hence, the total mass of the material accumulated on the equilibrium sphere always remains negligible in comparison to the stellar mass $M$ for physically motivated parameters. For the example parameters considered in Fig.~\ref{fig:RNKS_HLyt} we find ($v = 0.1\,c$, $b_{\rm crit} \approx 1700\,M$, $m_p$ is the~proton mass, $n_p$ is the number density of particles)
\begin{align}
\frac{M_{\rm shell}}{M} &= \pi b^2_{\rm crit} m_p n_p v \frac{\Delta T}{M}  \\
&\approx 9 \times 10^{-17} \left( \frac{\Delta T}{100\, \rm{s} } \right) \left(\frac{2 M_\odot}{M} \right) \left( \frac{n_p}{10^6 {\rm cm}^{-3}} \right) \, . \nonumber
\end{align}
The reference number density of $10^6\,{\rm cm}^{-3}$ overestimates a typical number density of molecular clouds.

%====================
\section{Discussion and summary}
%====================

We demonstrated the importance of the spacetime geometry for equilibria and dynamics of particles around luminous compact stars. In the presented examples of spherically symmetric spacetimes deviating from the Schwarzschild solution, the radius of the equilibrium sphere can be larger by as much as a~factor of 2 with respect to the Schwarzschild equilibrium at the same intrinsic luminosity. Apart from this effect constituting an interesting theoretical problem, there are potentially important applications of the presented results in the astrophysical context of phenomena involving strong radiation originating in the vicinity of a compact gravitating object. Understanding the location of the gravity-radiation equilibrium surface is relevant for the observational constraints on the neutron star equation of state from the type-I luminous thermonuclear photospheric radius expansion X-ray bursts \cite{Ozel2006, Galloway2008, Kim2021}, or from oscillations of the material suspended above the star surface by the strong radiation \cite{Bollimpalli2019}. In the former case, as an example, the ratio between observed near-Eddington fluxes at the peak of the X-ray burst and near its end (the touchdown phase) provides information on the gravitational redshift at the stellar surface \cite{Damen1990}, related to our parameter $\xi_R$. The relation between the inferred redshift and the $M/R$ quantity that can be subsequently used to constrain the neutron star equation of state is thus depending on the underlying spacetime geometry through the metric function $\xi(r)$. Similarly, for the measurement proposed in \cite{Bollimpalli2019}, the frequencies of the equilibrium sphere oscillations depend on the spacetime geometry and the related location of the equilibrium, affecting the inferred mass and radius of the neutron star.

On the one hand, relaxing the Schwarzschild spacetime model assumption results in another source of uncertainty, on the other hand, it may offer a way to test strong gravity. What is more, the dynamical properties of such systems could manifest observationally, e.g., increased velocity of the expanding atmosphere of a bursting neutron star would necessarily manifest as a shift in the observed X-ray spectrum. These problems have not received sufficient attention from the scientific community so far.

There are multiple research avenues that should be followed in order to generalize the presented results and make them more physically relevant. In particular, the test-particle approximation can be relaxed and more realistic cases of optically thin \cite{Wielgus2015b, Bollimpalli2017} and optically thick \cite{Wielgus2016} levitating (suspended in the radiation field) atmospheres can be considered. The questions regarding the impact of the rotation of the star and the role of magnetic fields are also important as subjects of further studies that will necessarily render the presented toy model more complicated. In particular, \citep{Wielgus2019} demonstrated that for a rotating source with a realistic radiation stress-energy tensor \citep{Miller1996} the radial equilibrium is only present in the star's equatorial plane, with radiation drag forces operating efficiently in the poloidal direction, destabilizing off-equatorial equilibria on relevant timescales for dimensionless spins as small as $a = 0.05$. For a fixed luminosity and other system parameters, the equatorial equilibrium occurs at a slightly larger radius if the star is rotating. Thus, a toroidal configuration, puffed-up by the pressure of the accumulated gas around the test-particle equatorial equilibrium is expected. Such two-dimensional equilibria have not been calculated so far (see \citep{Wielgus2015b,Wielgus2016} for a non-rotating spherically symmetric case).

\section*{Acknowledgements}
 We thank Marek Abramowicz and W\l odek Klu\'zniak for discussions. We also thank Alexandra Elbakyan for her contributions to the open science initiative. %MW acknowledges support from the European Research Council advanced grant “M2FINDERS - Mapping Magnetic Fields with INterferometry Down to Event hoRizon Scales” (GrantNo. 101018682).

%%%%%%%%%%%%%%%%%%%%%%%%%%%%%%%%%%%%%%%%%%%%%%%%%%



%%%%%\bibliographystyle{h-physrev_titulos}
%%%%%\bibliography{bibliography} % if your bibtex file is called example.bib

\label{lastpage}
\end{document}